\documentclass[review,authoryear,times]{elsarticle}
\journal{Radiation Physics and Chemistry (XAFS-2018)}

\usepackage{graphicx}% Include figure files

\usepackage{color}

\usepackage{amssymb}

\biboptions{authoryear}

\usepackage{hyperref}
\hypersetup{
	colorlinks   = true, %Colours links instead of ugly boxes
	urlcolor     = blue, %Colour for external hyperlinks
	linkcolor    = black, %Colour of internal links
	citecolor   = red %Colour of citations
}

\begin{document}

\begin{frontmatter}

\title{Treatment of disorder effects in X-ray absorption spectra beyond the conventional approach}

\author[ISSP,IROAST]{Alexei Kuzmin\corref{mycorrespondingauthor}}
\cortext[mycorrespondingauthor]{Corresponding author}
\ead{a.kuzmin@cfi.lu.lv}
\ead[url]{http://www.cfi.lu.lv}

\address[ISSP]{Institute of Solid State Physics, University of Latvia, Kengaraga street 8, LV-1063 Riga, Latvia}

\address[IROAST]{International Research Organization for Advanced Science and
	Technology (IROAST), Kumamoto University, 2-39-1 Kurokami, Chuo-ku, Kumamoto 860-8555, Japan}

\author[SB]{Janis Timoshenko}
\address[SB]{Department of Materials Science and Chemical Engineering, Stony Brook University, Stony Brook, NY 11794, USA}

\author[UP]{Aleksandr Kalinko}
\address[UP]{Universit\"{a}t Paderborn, Naturwissenschaftliche Fakult\"{a}t, Department Chemie, Warburger Strasse 100, 33098 Paderborn, Germany}

\author[ISSP]{Inga Jonane}

\author[ISSP]{Andris Anspoks}

\begin{abstract}
The contribution of static and thermal disorder is one of the largest challenges for the accurate determination of the atomic structure from the
extended X-ray absorption fine structure (EXAFS). 
Although there are a number of generally accepted approaches to solve this problem, 
which are widely used in the EXAFS data analysis, they often provide less accurate 
results when applied to outer coordination shells around the absorbing atom.
In this case, the advanced techniques based on the molecular dynamics and reverse 
Monte Carlo simulations are known to be more appropriate: their strengths and 
weaknesses are reviewed here. 
\end{abstract}


\end{frontmatter}

%\linenumbers

%\newpage

\section{Introduction}
\label{intro}

X-ray absorption spectroscopy (XAS) is an excellent tool to probe the local environment in crystalline, nanocrystalline and disordered solids, liquids and gases in a wide range 
of {\it in situ} and {\it in operando} conditions %(\cite{Ortega2012,Bordiga2013,Kuzmin2014exafs,Oversteeg2017,Mino2018})
(\cite{Oversteeg2017,Mino2018}). 
With the increased availability of synchrotron radiation sources and the tremendous  improvement in their parameters, the popularity of the technique has increased, and the quality of the experimental X-ray absorption spectra has improved significantly. 
As a result, more accurate and reliable structural information can be extracted 
from the extended X-ray absorption fine structure (EXAFS) located above the absorption 
edge of an element. 

The quantitative analysis of EXAFS became possible due to significant advancements 
in the theory (\cite{Rehr2000,Natoli2003,Rehr2009}), however, accurate treatment of  disorder effects is still the biggest difficulty. 
The problem becomes especially acute when it comes to the outer coordination 
shells around the absorbing atom, where the overlap of the shells and the effect of the disorder are mixed with the multiple-scattering (MS) contributions.    

This paper reviews the existing approaches commonly used to solve the problem of disorder 
in EXAFS and discusses the strengths and weaknesses of two advanced techniques based 
on the molecular dynamics  and reverse Monte Carlo  methods.

\section{Conventional approach to disorder in EXAFS}
\label{conv}

In this section, we will briefly summarize different conventional approaches to the treatment of disorder in EXAFS.  

The X-ray absorption coefficient $\mu(E)$ in the one-electron approximation is proportional 
to the transition rate between the initial core-state $i$ and the final excited-state $f$ 
of an electron, which is given by the Fermi’s Golden rule 
\begin{equation}
  \mu(E) \propto \sum_f \left | \langle f| \hat{H} |i \rangle \right |^2 \delta (E_f - E_i - E)
\label{eq1}  
\end{equation}
where $E=\hbar \omega$ is the X-ray photon energy, and the transition operator 
$\hat{H} = \hat{\epsilon} \cdot \vec{r}$ in the dipole approximation. Note that the final state of the electron is the relaxed excited state 
in the presence of the core-hole screened by other electrons. 

The characteristic time of the photoabsorption process (\ref{eq1}) is about 10$^{-15}$--10$^{-16}$~s and is determined by several processes: the transition time between initial ($i$) and final ($f$) states, the core-hole lifetime,  the excited photoelectron relaxation time and the lifetime of the photoelectron out of atom related to its mean-free path (MFP). Note that this time is significantly shorter than
the characteristic time ($\sim$10$^{-13}$--10$^{-14}$~s) of thermal vibrations. Therefore,
atoms can be considered as frozen at their instantaneous positions during the excitation process,
and the experimental X-ray absorption spectrum corresponds to the average over all atomic configurations during the time of experiment. 

The  oscillating part of the absorption coefficient $\chi^l(E)$ located above the absorption edge of orbital type $l$ is defined as 
\begin{equation}
\chi^l(k)= (\mu(E)-\mu_0(E)-\mu_b(E))/\mu_0(E))
\label{eq2}
\end{equation}
where $\mu_b(E)$ is the background absorption, and $\mu_0(E)$ is the atomic-like absorption due to an isolated absorbing atom (\cite{Lee1981}).
The  wave number $k$ of the excited photoelectron is related to its kinetic energy $(E-E_0)$ by
$k=\sqrt{(2m_e/\hbar^2)(E-E_0)}$, where $m_e$ is the electron mass, $\hbar$ is the Plank's constant,
and $E_0$ is the threshold energy, i.e., the energy of a free electron with zero momentum.

Within the framework of MS theory, EXAFS $\chi^l(k)$ is described  using a series 
\begin{eqnarray}
\chi^l(k) &=& \sum_{n=2}^{\infty} \chi^l_n(k), \nonumber \\
\chi^l_n(k)  &=& \sum_{j}A_n^l(k,R_j) \sin[2kR_j+\phi_n^l(k,R_j)]  
\label{eq3}
\end{eqnarray}
which includes contributions $\chi^l_n(k)$  from the $(n-1)$-order
scattering processes of the excited photoelectron by the
neighbouring atoms, before it returns to the absorbing atom (\cite{RuizLopez1988,Rehr2000}). 
The fast convergence of the MS series occurs at least 
at high-$k$ values due to the finite lifetime of the excitation, 
the scattering path lengths, interference cancellation effects and path disorder. In practice, the MS contributions up to 
the 8th-order can be calculated using {\it ab initio} FEFF code (\cite{FEFF8,FEFF9}).

An alternative description of the EXAFS $\chi^l(k)$  in terms of the $n$-order
distribution functions $g_n(R)$  is also known
\begin{eqnarray}
\chi^l(k) &=& \int 4\pi R^2 \rho_0 g_2(R)
[\chi_2^{oio}(k) + 
%\chi_4^{oioio}(k) 
 +  \ldots] dR \nonumber \\
 & + &  \int\!\!\!\int\!\!\!\int 8 \pi^2 R_1^2 R_2^2 \sin(\theta)
\rho_0^2 g_3(R_1, R_2, \theta) \nonumber \\
 &\times& [2\chi_3^{oijo}(k) +
2\chi_4^{oiojo}(k) + 
%\chi_4^{oijio}(k) +  \chi_4^{ojijo}(k) +
 \ldots ] dR_1 dR_2 d\theta  \nonumber \\
% &+&  
% \int\!\!\!\int\!\!\!\int\!\!\!\int\!\!\!\int 8 \pi^2 R_1^2 R_2^2 R_3^2 \sin[\theta]
% \rho_0^3  g_4(R_1, R_2, \theta, R_3, \Omega)  \nonumber \\
% & \times& (2\chi_4^{oijko}(k) + 2\chi_4^{oikjo}(k) +
% 2\chi_4^{ojiko}(k)  + \ldots )
% dR_1 dR_2 d \theta dR_3 d \Omega  \nonumber \\
 & +&  \ldots
\label{eq4}
\end{eqnarray}
where $\rho_0$\ is the average density of a system and $\chi_m(k)$
are the MS EXAFS signals of the $(m-1)$\ order
generated within a group of atoms (o, i, j, \ldots) described by
$g_n$ (\cite{Filipponi1995a,Filipponi1995b}).
%(\cite{Filipponi1991,Filipponi1995a,Filipponi1995b}). 
This approach was realized in the GNXAS code (\cite{DiCicco1995,GNXAS}), which 
is able to account for the two-body ($g_2$), three-body ($g_3$) and four-body ($g_4$) distribution functions.

The analytical expression for EXAFS can be greatly simplified 
when one needs to extract information only on the first coordination shell of the absorbing atom.

The contribution of the first coordination shell to the total EXAFS spectrum can be usually 
isolated  by Fourier filtering procedure and analysed within the single-scattering approximation, since the length of all MS paths is longer than the first coordination shell radius. 
Thus, only the first term of the series given by Eq.~(\ref{eq3}) 
remains. In the case of a Gaussian distribution (or in the harmonic approximation), 
the EXAFS expression takes a simple form 
\begin{eqnarray}
\chi^l_2(k) & = & S_0^2 \sum_{i} N_i \frac{|f^l_{\rm eff}(k,R_i)|}{kR_i^2} 
\exp\left[-\frac{2 R_i}{\lambda(k)}\right] \nonumber \\
 & \times &  \sin[2kR_i+\phi^l(k,R_i)]  \exp(-2\sigma_i^2 k^2)
\label{eq5}
\end{eqnarray}
where $S_0^2$ is a scaling factor; $N_i$ is the coordination number; 
$R_i$ is the interatomic distance; $\lambda(k)$ is the photoelectron MFP; 
$f^l_{\rm eff}(k,R)$ and $\phi^l(k,R)$ are the photoelectron effective scattering amplitude and phase shift
functions  (\cite{Sayers1971,Lee1975}). The sum in Eq.~(\ref{eq5}) is taken over groups of atoms located at different distances from the absorber.

For moderate disorder, when distribution  of interatomic distances becomes asymmetric, the EXAFS equation can be expressed using the cumulant decomposition (\cite{Bunker1983,Dalba1993}).  
%as
%\begin{eqnarray}
%\chi^l_2(k) & = & S_0^2  \sum_{i}  N_i \frac{|f^l_{\rm eff}(k,R_i)|}{k R^2_i}
%  \exp\left[-\frac{2 R_i}{\lambda(k)}\right]
%\nonumber \\
%& \times &  \sin[2kR_i - {4 \over 3} C_{3i} k^3 + \phi^l(k,R_i)] \nonumber \\
%& \times & \exp(-2 \sigma_i^2 k^2 + {2 \over 3} C_{4i} k^4 ),
%\label{eq6}
%\end{eqnarray}
%where  $C_{3i}$ and $C_{4i}$ are cumulants of the effective distribution %(\cite{Fornasini2001}). 
The cumulant model is often useful for the analysis of anharmonic and thermal expansion effects  (\cite{Tranquada1983,Fornasini2017}), nanoparticles
 (\cite{Clausen2000,Sun2017}) and  disordered materials (\cite{Dalba1995age,Okamoto2002}).

Sometimes, the first coordination shell around the photoabsorber is so strongly distorted that the cumulant series  
%given by Eq.~(\ref{eq6}) 
does not converge.
In this case, the EXAFS formula expressed in terms of the radial distribution function (RDF) $G(R)$  
\begin{eqnarray}
\chi^l_2(k) &=& S_0^2 \int_{R_{\rm {min}}}^{R_{\rm {max}}} G(R) \frac{|f^l_{\rm eff}(k,R)|}{k R^2}
 \nonumber \\
& \times&  \sin[2kR  + \phi^l(k,R_i)]  \exp\left[-\frac{2 R_i}{\lambda(k)}\right] dR
\label{eq7}
\end{eqnarray}
should be used instead (\cite{Stern1975,Lee1981}). The RDF $G(R)$ 
defines the probability of finding an atom in a spherical shell $dR$ at the distance $R$ from the photoabsorber.
The number $N$\ of atoms located in the range between
$R_{\rm {min}}$\ and $R_{\rm {max}}$\  is given by the integral $N =
\int_{R_{\rm {min}}}^{R_{\rm {max}}} G(R) dR$. To determine RDF
$G(R)$ from Eq.~(\ref{eq7}), the regularization technique (\cite{Babanov1981,Ershov1981,Kuzmin2000}) 
can be used to solve this integral equation as an ill-posed problem 
without any preliminary assumption on the shape of the RDF. This approach 
was recently used to reconstruct the local structure in several tungstates 
MWO$_4$ (M=Ni, Cu, Zn and Sn) (\cite{Kalinko2011,Anspoks2014b,Kuzmin2015})
and in molybdate CuMoO$_4$ (\cite{Jonane2018cmo}), where the Jahn-Teller effect is responsible for a strong distortion of structural units. 
It was demonstrated recently that the RDF $G(R)$ of atoms can be reliably extracted 
up to distant coordination shells using neural network approach (\cite{Timoshenko2018ann}).

\begin{figure}[t]
	\centering
	\includegraphics[width=0.75\textwidth]{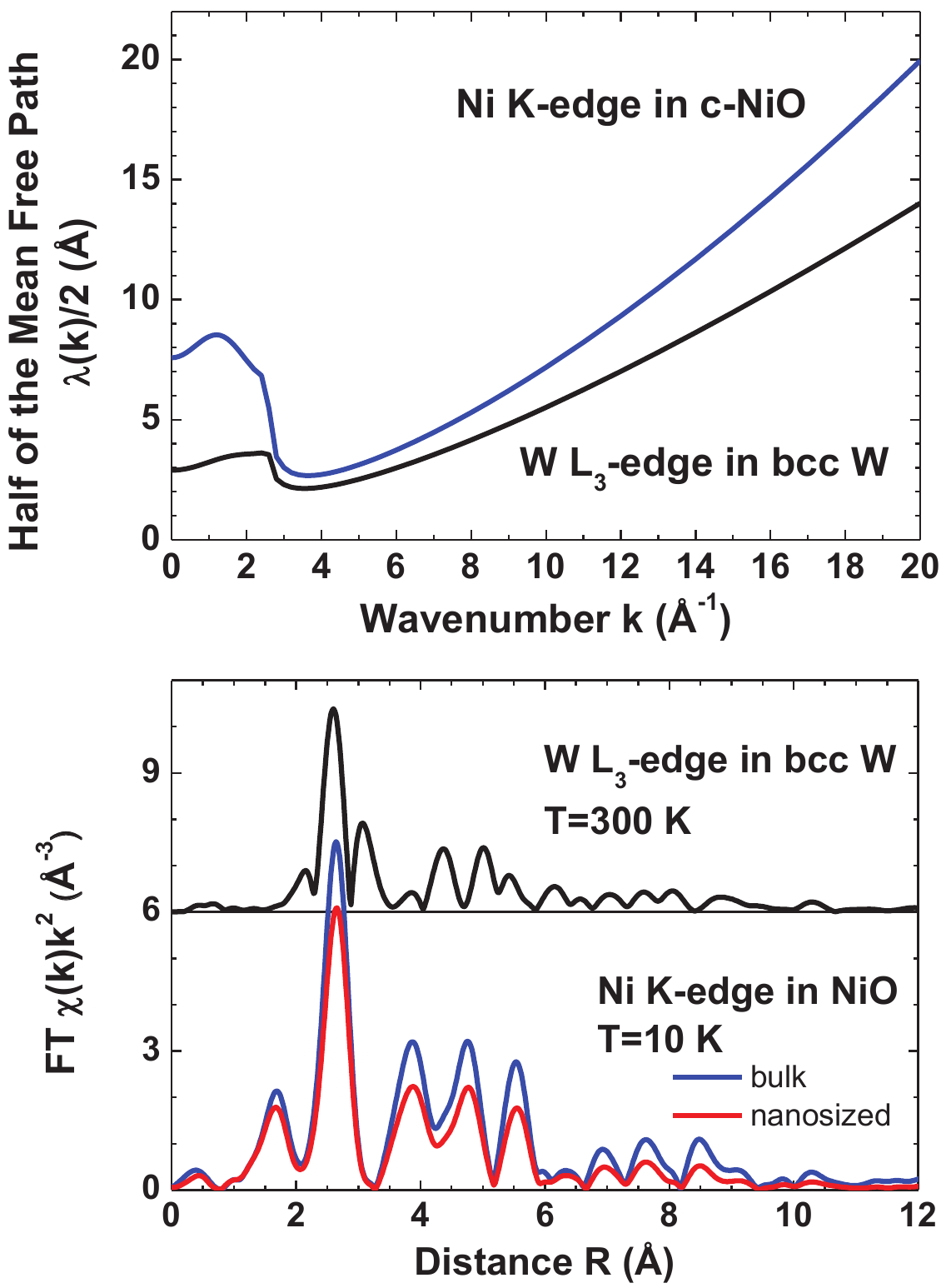}
	\caption{ Upper panel: Calculated photoelectron mean free path (MFP) $\lambda(k)$ for c-NiO and bcc W. Lower panel: 
	Fourier transforms of the experimental W L$_3$-edge and Ni K-edge EXAFS spectra $\chi(k)k^2$ in 
	bulk and nanosized NiO at $T=10$~K and in bcc tungsten at $T=300$~K, respectively. }
	\label{fig1}
\end{figure}

In crystalline and nanocrystalline materials, the experimental EXAFS spectrum often 
contains a significant amount of structural information on outer coordination shells, which is 
challenging to extract. 
It is possible to estimate the region of a structure around the absorber, which can potentially contribute into EXAFS,  from the photoelectron MFP.
Examples for bulk and nanocrystalline nickel oxide (\cite{Anspoks2012}) and  body-centred-cubic (bcc) tungsten (\cite{Jonane2018})
 are shown in Fig.~\ref{fig1}.  A half of the MFP $\lambda(k)$
gives an estimate of how far the excited photoelectron can propagate to be able to return back to the absorbing atom. 
The MFP $\lambda(k)$ depends strongly on the photoelectron wavenumber $k$ and increases at large $k$-values.  It is equal to about 10--20~\AA\ for NiO or bcc W 
at $k \approx 16$--20~\AA$^{-1}$. This means that when high-quality experimental EXAFS data are available in  large $k$-space range, one can expect to see structural contributions  from  atoms located in distant coordination shells. 
For example, the structural peaks in Fourier transforms 
of EXAFS can be recognized up to about 11~\AA\ in Fig.~\ref{fig1}
for bulk and nanosized NiO at $T=10$~K and for bcc W at $T=300$~K.

The possibility to analyse contributions from distant coordination shells is useful since 
it provides access to additional structural information. However, such analysis based on 
the conventional approaches faces a number of problems even for crystalline materials with a known structure, in which at least the mean-square relative displacement (MSRD) factors are variable model parameters.  

\begin{figure}[t]
	\centering
	\includegraphics[width=0.75\textwidth]{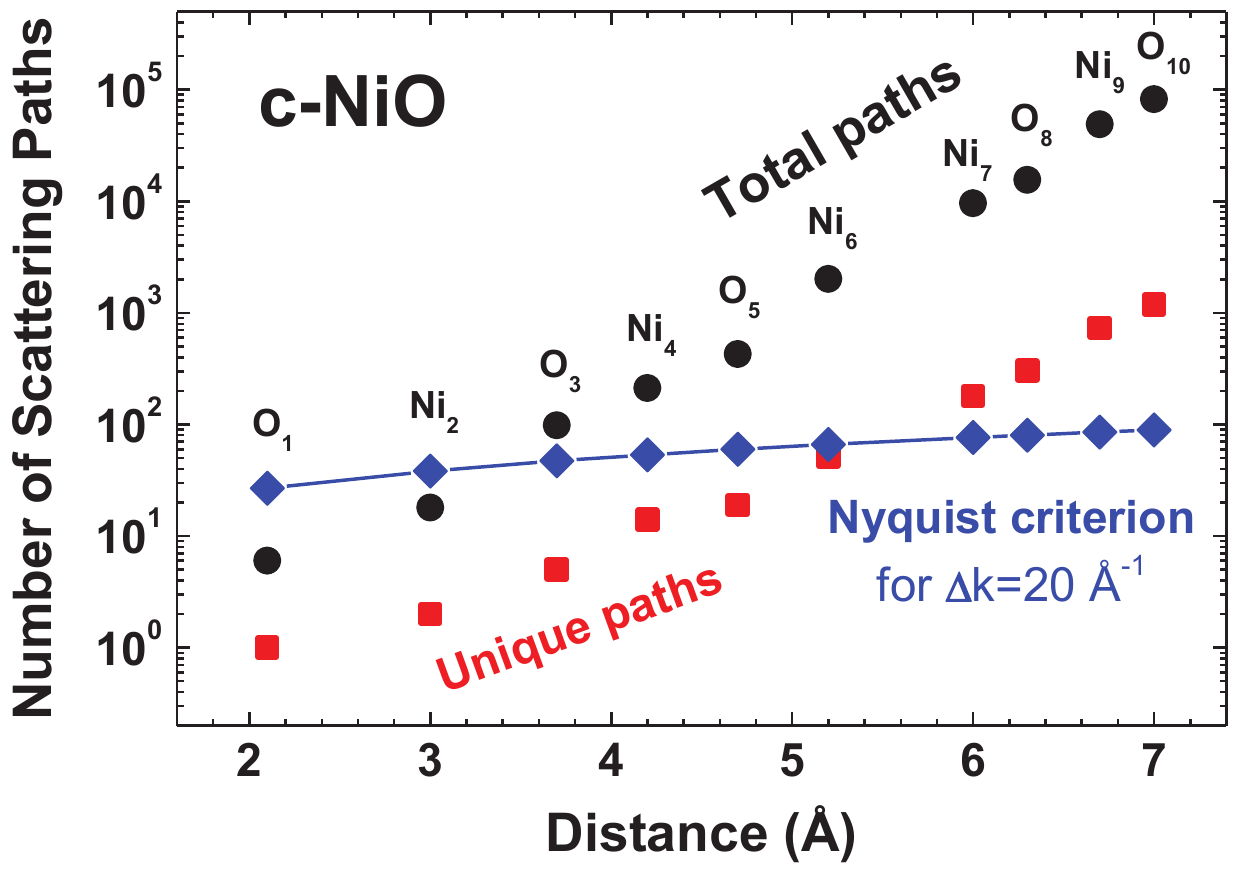}
	\caption{ The dependence of the number of scattering paths on the cluster size around the absorbing nickel atom in NiO. }
	\label{fig2}
\end{figure}

The main problem is related to the number of model parameters, 
which increases exponentially when more coordination shells are included to the model  (\cite{Kuzmin2014exafs}). For example, in the case of bulk NiO with a
rock-salt structure, the total number of scattering paths, 
the number of unique paths due to the cubic symmetry and the maximum number 
of fitting parameters, which can be used in the EXAFS model according to the Nyquist
criterion ($N_{\rm par} = 2 \Delta k \Delta R / \pi $) evaluated for relatively long EXAFS signal with $\Delta
k$=20~\AA$^{-1}$, are shown in Fig.~\ref{fig2} as a function of 
the cluster radius $R$ around the photoabsorbing nickel atom. 
Note that the Nyquist criterion is not satisfied above $R$$\sim$5.5~\AA, 
when cubic crystal symmetry is taken into account, but this
distance decreases significantly down to $R$$\sim$3.5~\AA\ 
in a nanomaterial.

To reduce the number of model parameters, one can evaluate the MSRD factors semiempirically from correlated Einstein or Debye models (\cite{Sevillano1979,Vaccari2006}), but again different  Einstein or Debye  temperatures are required for each MS path. Besides, 
these models of lattice dynamics ignore anisotropy of the phonon spectra.

Another approach is to calculate MSRD parameters from the phonon projected density of states using the Debye integral
\begin{equation}
\sigma^2_R(T) = \frac{\hslash}{2\mu_R} \int_0^\infty \frac{1}{\omega} \coth \left( \frac{\hslash \omega}{2 k_B T} \right)  \rho_R(\omega) d\omega
\label{eq8}  
\end{equation}
where $\mu$ is the reduced mass associated with the MS path, and $k_B$ is the Boltzmann's constant.  The vibrational density of states $\rho_R(\omega)$ projected on $R$
can be obtained from first-principles calculations of the dynamical matrix of force constants  (\cite{Vila2007,Rehr2009,FEFF9}).
However, this approach uses (quasi-)harmonic approximation, requires {\it a priori} knowledge of structure and can be computationally expensive. 

An alternative solution which allows one to account simultaneously for the MS contributions 
and disorder effects is to rely on  atomistic simulations such as the 
molecular dynamics (MD)  and reverse Monte-Carlo  (RMC) methods combined with ab initio MS calculations.

\begin{figure*}[t]
	\centering
	\includegraphics[width=1\textwidth]{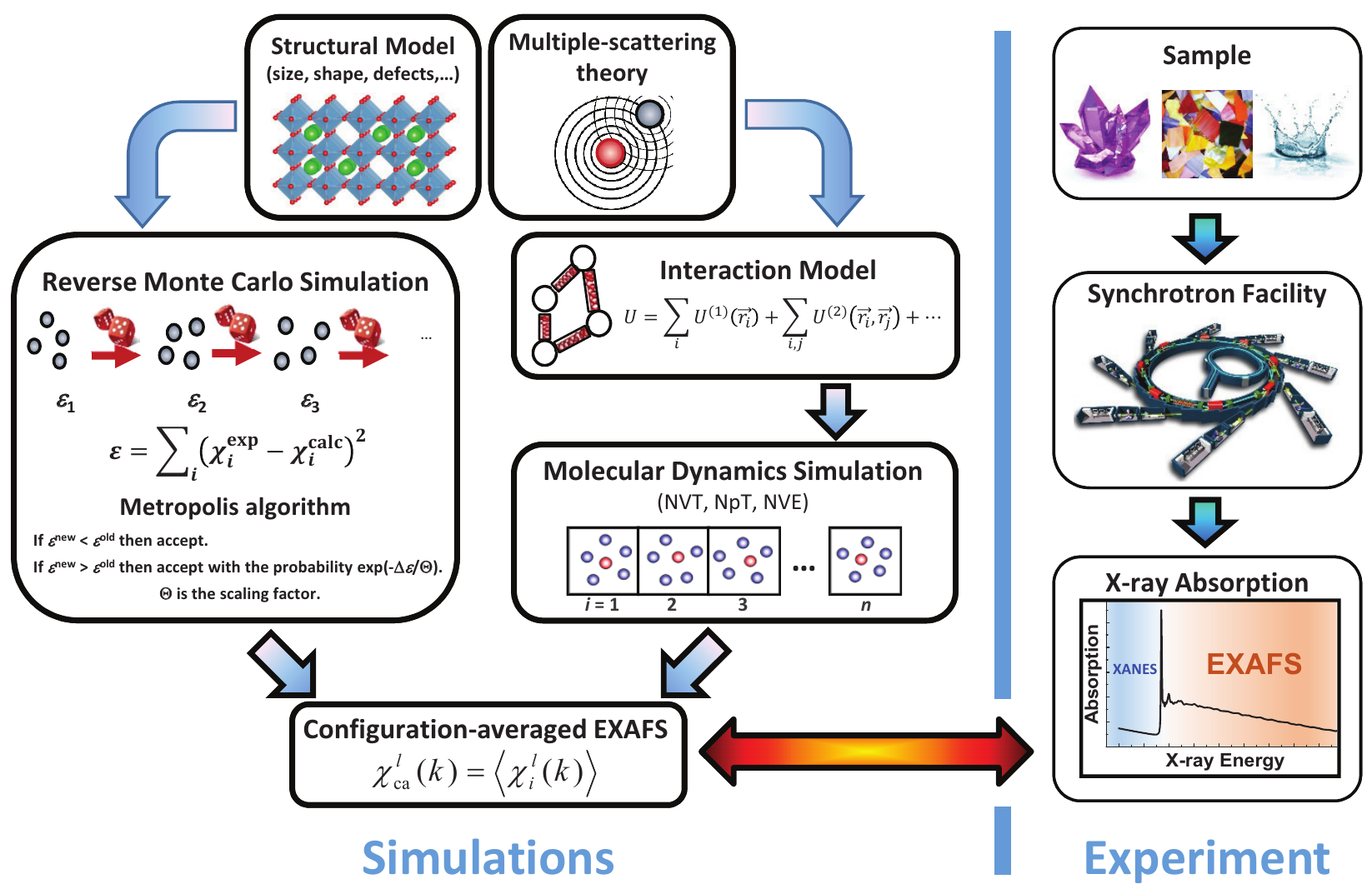}
	\caption{Scheme of EXAFS analysis using reverse Monte Carlo and molecular dynamics methods. }
	\label{fig3}
\end{figure*}

\section{Atomistic simulations of EXAFS}
\label{simul}

MD (\cite{Alder1957}) and RMC (\cite{McGreevy1988}) methods are known for 
a long time, however their application in the field of X-ray absorption spectroscopy is still  scarce. 
%rather sparse
The use of both methods requires significant computing resources, 
so their development has been directly related to the advances in computer technologies.  

The first use of MD simulations to reproduce the experimental EXAFS is dated back to 
the middle of nineties, when the method was applied to study the hydration  
of ions in aqueous solutions (\cite{Angelo1994,Angelo1996,Palmer1996,Kuzmin1997zn}). 
The advantages of the RMC method were realized even earlier at the beginning of nineties,
when it was used to interpret EXAFS of amorphous Si and crystalline AgBr  (\cite{Gurman1990rmc}),
liquid KPb alloys (\cite{Bras1994}) and superionic glasses (\cite{Wicks1995}).

There are several common features for the MD and RMC methods.
The simulation result is represented as one or more atomic configurations (``snapshots''), suitable to generate the configuration-averaged (CA)
EXAFS, which includes static and dynamic disorder and can be directly compared to experimentally measured EXAFS.  
The static disorder is due to a number of 
different atomic dispositions, corresponding to minima of the potential energy surface.
Examples of systems with the static disorder include non-crystalline materials such as glasses, amorphous solids and liquids, nanocrystals and thin films with atomic structure relaxed due to the size or thickness reduction effect, and
materials with structural defects (e.g., vacancies or grain boundaries).  
Dynamic disorder arises from temperature-dependent fluctuations in the atomic positions from the equilibrium structure. 

The  CA EXAFS spectra for different absorption edges 
can be calculated from the same set of atomic coordinates and used in the analysis, thus 
improving the reliability of the structural model (\cite{Timoshenko2014}).   
During a simulation, the atoms are placed in a cell of the required size and shape, 
often with periodic boundary conditions (PBC) in order to avoid effects associated with the surface. 
Note that using PBC limits the maximum cluster radius, for which EXAFS calculations 
can be safely performed to avoid artificial  correlation effects, to half the minimum cell size. 
There are also two non-structural parameters, $\Delta E_0$ and $S_0^2$, 
which are required for comparison with the experimental EXAFS. They can be determined from 
the analysis of reference materials or obtained by best matching the experimental and calculated EXAFS spectra.  

The scheme of the MD and RMC methods is shown in Fig.~\ref{fig3}. The structural model of a material 
is constructed first in both cases, and the {\it ab initio} MS code, such as FEFF (\cite{FEFF8}) or GNXAS (\cite{GNXAS}),
is used to calculate EXAFS for each atomic configuration during the simulation. 

The principal difference between two methods is that no fitting of experimental EXAFS is performed in the MD-EXAFS approach, and the structure obtained in the MD simulation is used ``as-is'' for the calculation of the CA EXAFS. Note that the number of required atomic configurations  and the time step between them should be carefully estimated for each particular case to obtain the proper CA signal. 
On the contrary, the structural model is modified at each RMC iteration to minimize the difference between the experimental and CA EXAFS in the RMC-EXAFS approach. 

To perform MD simulations, a model of interactions between atoms is required. 
In classical MD (CMD), the empirical interatomic potential is employed, that 
significantly reduces the requirements for computing resources. Besides, 
the MD-EXAFS approach is suitable for a validation of interatomic potential 
along with other conventionally employed properties of a material (\cite{DiCicco2002pb,Kuzmin2009,Kuzmin2016,Bocharov2017}).
{\it Ab initio} MD (AIMD) based on density functional theory (DFT) formalism
is also accessible nowadays but is extremely computationally
expensive. It is important that in the MD simulation, initial model of the atomic structure
is evolving in time within one of the canonical (NVT), isothermal–isobaric (NpT) 
or microcanonical (NVE) ensembles following to classical Newtonian laws of
motion both in CMD and AIMD. Therefore, such simulations 
cannot  be used to model the motion of atoms at low temperatures, where
the zero-point oscillations of atoms play an important role (\cite{Yang2012}).
In this case, instead, more complex methods should be used, such as, for example, the path-integral MD (\cite{Dominik1996}).

Note that recent developments of X-ray free-electron laser (X-FEL) facilities open new possibilities to probe the ultrafast excited state dynamics using X‐ray absorption spectroscopy (\cite{Lemke2017}). Such experiments provide information 
on the femtosecond nuclear wavepacket dynamics, which can be described by 
first-principles quantum dynamics simulations (\cite{Capano2015}).

The MD simulations can be performed, for example, either by one of the CMD codes as LAMMPS (\cite{Plimpton1995}), GULP (\cite{Gale2003}) or DL\_POLY (\cite{Todorov2006}), or using AIMD codes  as CP2K (\cite{VandeVondele2005cp2k}), VASP (\cite{Kresse1996}) or SIESTA (\cite{Soler2002siesta}).  After accumulating  
the required number of atomic configurations, one can employ, for example, the EDACA code (\cite{Kuzmin2009,Kuzmin2016}) to generate the CA EXAFS spectrum. 

In RMC simulation, the position of atoms in the configuration is usually 
randomly modified at each iteration, and the CA EXAFS signal is calculated. 
The decision to accept or reject the new atomic configuration is made based on the Metropolis algorithm (\cite{Metropolis1953}), taking into account the difference (residual) between the experimental  and simulated  data in either $k$ or $R$ space, or simultaneously in $k$ and $R$-spaces using the wavelet transformation  (\cite{Timoshenko2009wt}). 
At this point, various chemical or geometrical constraints can be 
easily implemented, by assigning some penalty to the residual value. 
For example, one can avoid situations when the atoms are getting too close or 
too far from each other, when non-physical values of some bond
angle are found (\cite{Tucker2007rmcprofile}), or when the coordination number
for some atom deviates from the expected one (\cite{McGreevy2001}), etc.
The  efficiency of the RMC process can be significantly improved by  using an 
evolutionary algorithm (EA)  together with a simulated annealing scheme   
(\cite{Timoshenko2012rmc,Timoshenko2014rmc}).
The RMC method relies on stochastic process, so it will generate different 
final sets of atomic coordinates upon restarting simulation several times 
from different starting conditions. However, it is expected that the results will be statistically close in terms of the distribution functions.  
Note that RMC method tends to converge to the most disordered solution consistent with 
the experimental data (\cite{Tucker2007rmcprofile}).

Some of the software packages for RMC-EXAFS simulations include RMC-GNXAS (\cite{DiCicco2005}), RMCProfile (\cite{Tucker2007rmcprofile}), 
EPSR-RMC (\cite{Bowron2008epsr}), SpecSwap-RMC (\cite{Leetmaa2010specswap}),
RMC++/RMC\_POT (\cite{Gereben2007,Gereben2012rmc_pot}) and EvAX (\cite{Timoshenko2014rmc}).

Note that in addition to the MD and RMC methods, the average atomic configuration required 
to compute CA EXAFS can also be generated from a Monte Carlo simulation based on  
interatomic potentials  (\cite{Hansen1997,CancheTello2014,House2017}) 
or atomic displacement parameters obtained from lattice dynamics calculations
(\cite{Duan2016,Lapp2018}). 

\section{Examples of MD/RMC-EXAFS applications}
\label{examples}

In this section the specific capabilities of the MD-EXAFS and RMC-EXAFS methods will be demonstrated.

\begin{figure}[t]
	\centering
	\includegraphics[width=0.75\textwidth]{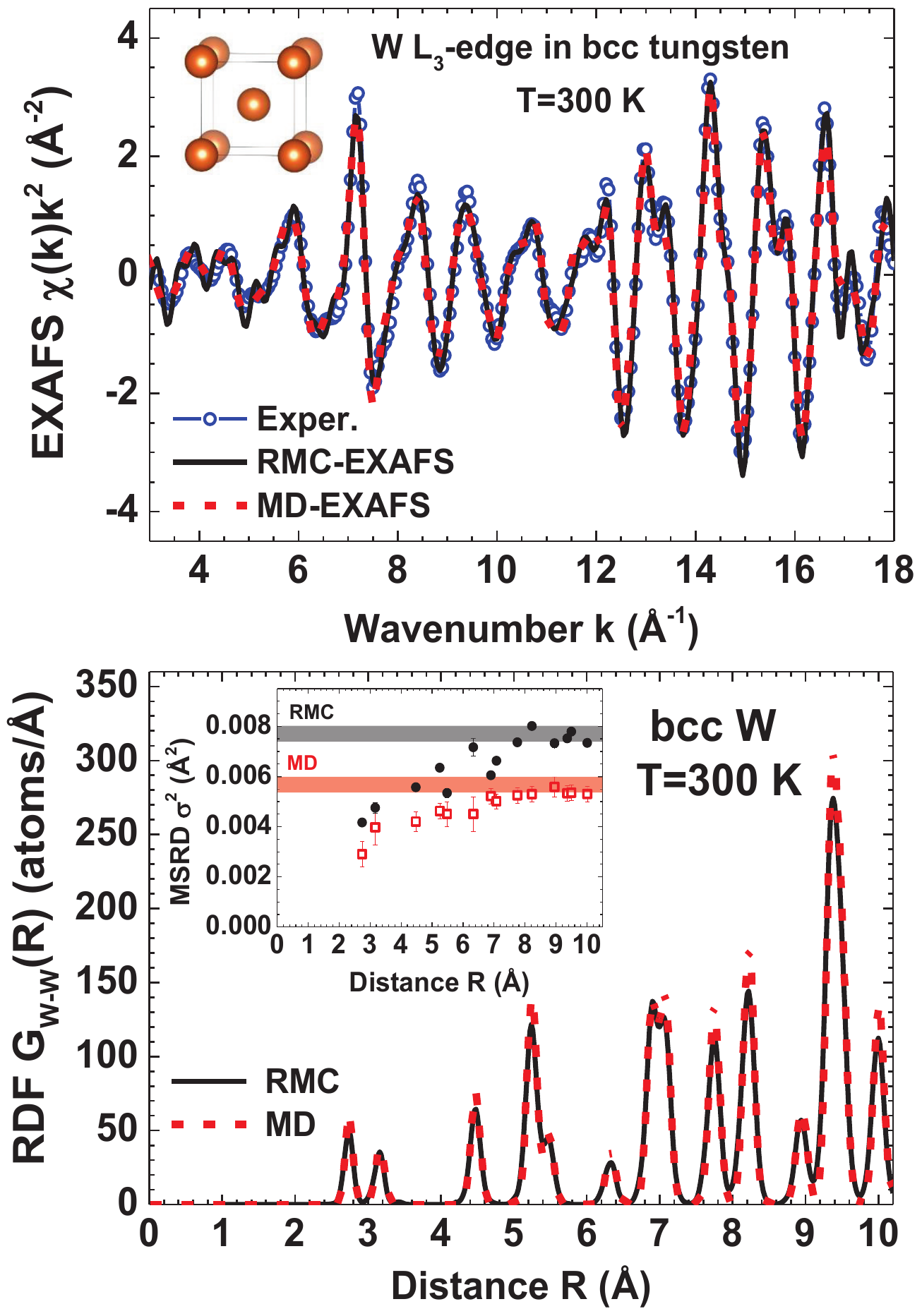}
	\caption{ Upper panel: Comparison of the experimental and 
		calculated by the RMC and MD methods W L$_3$-edge EXAFS $\chi(k)k^2$ 
		of bcc W at $T=300$~K. 
		Lower panel: Radial distribution functions (RDF) $G_{\rm W-W}(R)$ 
		obtained by RMC and MD simulations. Inset: Dependence of the MSRD $\sigma^2_{\rm W-W}(R)$ on distance. Two horizontal lines correspond to a sum of two MSDs of tungsten. }
	\label{fig4}
\end{figure}

The first example is concerned with the lattice dynamics in bcc tungsten (\cite{Jonane2018}). 
High-quality experimental W L$_3$-edge EXAFS spectrum was recorded at $T=300$~K up to $k=18$~\AA$^{-1}$ (Fig.~\ref{fig4} (upper panel)) and includes contributions from the coordination shells with a radius of at least up to $\sim$11~\AA\ (Fig.~\ref{fig1} (lower panel)). 
% It can be well reproduced using both MD and RMC methods. 
The NVT MD simulations were performed by the GULP code (\cite{Gale2003}) using a supercell of 7$a_0$$\times$7$a_0$$\times$7$a_0$ size ($a_0=3.165$~\AA) and 
a time step of 0.5~fs. The interactions were described by the second nearest-neighbour 
modified embedded atom method (2NN-MEAM) potential (\cite{Lee2001}). After equilibration 
during 20~ps, the atomic configurations were accumulated during the production run of 20~ps and used to calculate the CA EXAFS. 
The RMC/EA calculations were performed by the EvAX code (\cite{Timoshenko2014rmc}) using a supercell of 5$a_0$$\times$5$a_0$$\times$5$a_0$ size
to get best possible agreement between the Morlet wavelet transforms (WTs)
of the experimental and calculated EXAFS spectra. 
Good agreement with the experimental EXAFS data was obtained for both MD-EXAFS and RMC-EXAFS 
approaches (Fig.~\ref{fig4} (upper panel)). Next, the atomic configurations were 
used to calculate the RDFs $G_{\rm W-W}(R)$ and the radial dependence of the MSRD factors 
$\sigma^2(R)$. At long distances, when correlation in atomic motion becomes negligible, the MSRD  $\sigma^2_{\rm W-W} = 2\sigma^2_{\rm W}$ (see the inset in Fig.~\ref{fig4} (lower panel)).
The obtained  mean square displacements  (MSD)  $\sigma^2_{\rm W}$ 
are in agreement with previously reported experimental and theoretical results (\cite{Jonane2018}). 
Thus, the analysis of distant coordination shells allows extracting information 
on the MSD of atoms,  which otherwise requires a diffraction experiment.

\begin{figure}[t]
	\centering
	\includegraphics[width=0.75\textwidth]{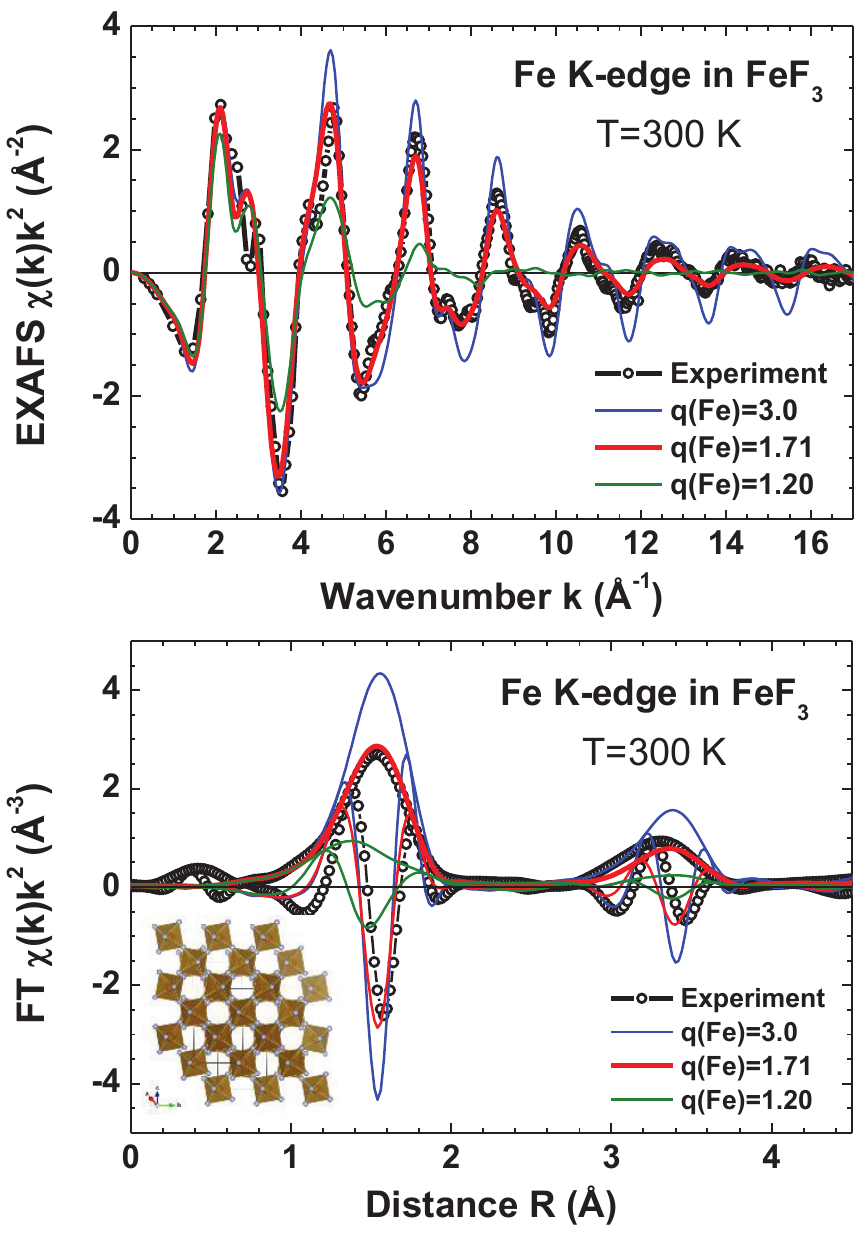}
	\caption{ Comparison of the experimental and calculated Fe
		K-edge MD-EXAFS $\chi(k)k^2$ spectra and their Fourier transforms
		(FTs) (modulus and imaginary parts are shown) in FeF$_3$ at
		$T$=300~K. Only few spectra calculated for different effective iron
		charges are shown for clarity.}
	\label{fig5}
\end{figure}

In the second example, the use of the MD-EXAFS approach for the validation 
of the interatomic potential model is shown on the example of iron fluoride (FeF$_3$) (\cite{Jonane2016}). The crystalline lattice of rhombohedral FeF$_3$ is composed of
FeF$_6$ octahedra joined by corners with the bond angle Fe--F--Fe between two adjacent octahedra equal to $\sim$153$^\circ$. The MD simulations were carried out using
a simple empirical potential, including two-body (Fe--F and F--F)
and three-body (Fe--F--Fe) interactions. It was found that different sets of the optimized potential parameters, corresponding to the iron effective charge $q$(Fe) in the 
range of 1.2--3.0, reproduce equally well the static crystallographic structure 
of FeF$_3$.  This ambiguity was resolved by performing NVT CMD simulations and 
calculating the CA Fe K-edge EXAFS spectra (Fig.~\ref{fig5}). 
Strong sensitivity of EXAFS to the strength of the Coulomb interactions was found, 
thus allowing one to select the iron effective charge $q$(Fe)=1.71 giving 
the best overall agreement between the experimental and CA EXAFS spectra.

\begin{figure}[t]
	\centering
	\includegraphics[width=0.75\textwidth]{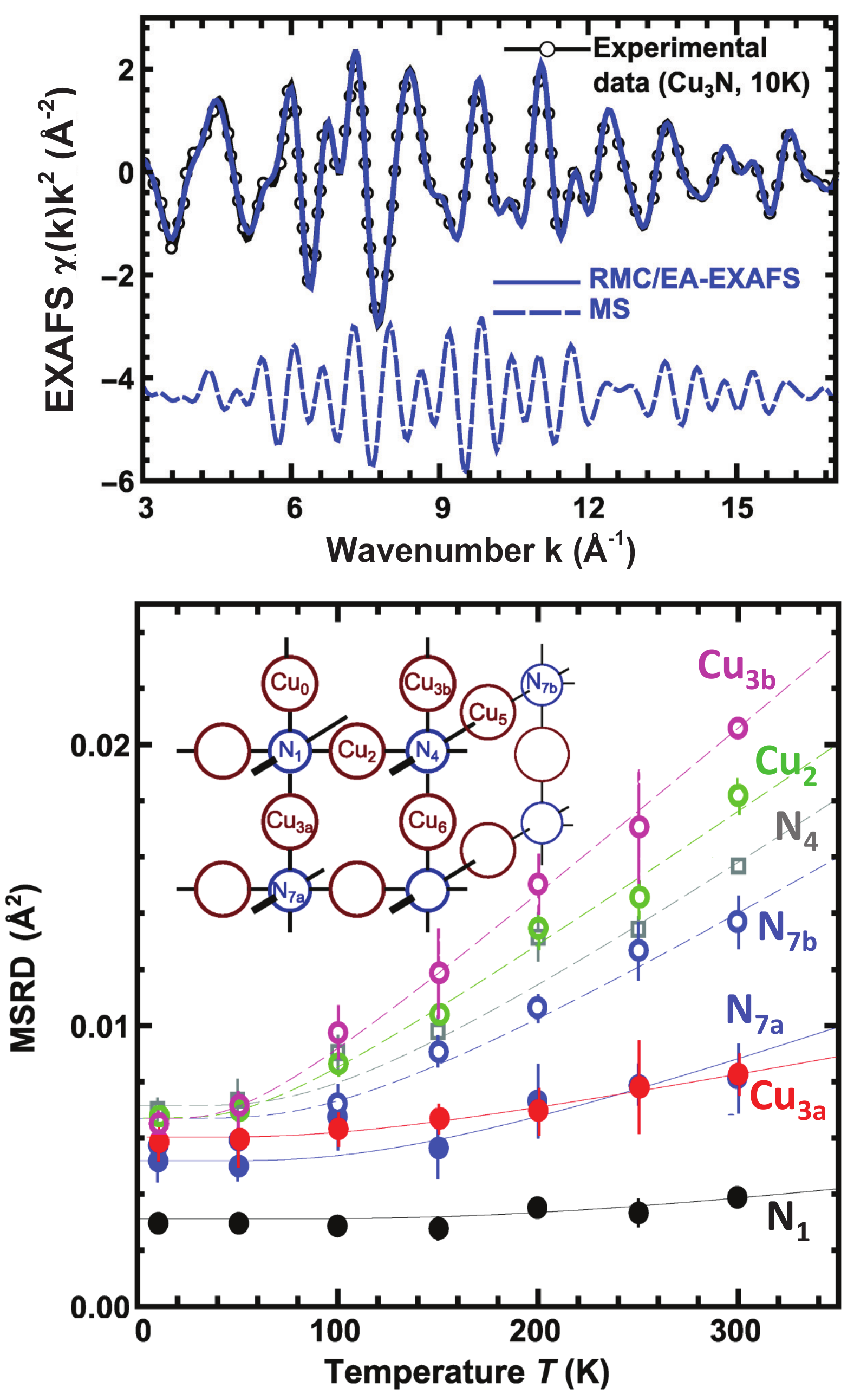}
	\caption{Upper panel: Comparison of the experimental and calculated CA Cu K-edge 
		EXAFS spectra of Cu$_3$N at $T$=10~K. Dashed line shows the total MS contribution. 
	Lower panel: Temperature dependencies of the MSRD factors for selected 
	Cu--N and Cu--Cu atom pairs in Cu$_3$N. Inset: The fragment of the Cu$_3$N structure.
	Coordination shells around the absorber Cu$_0$  are labelled. 
     }
	\label{fig6}
\end{figure}

Final example demonstrates the possibility to probe anisotropy and correlation of 
atomic motion in copper nitride (Cu$_3$N) using the RMC-EXAFS approach (Fig.~\ref{fig6}) (\cite{Timoshenko2017c}). 
Cu$_3$N has a unique cubic anti-perovskite-type structure (AB$_3$X), 
composed of NCu$_6$ octahedra joined by the corners with the A sites being vacant. 
High symmetry of its lattice is responsible for strong overlap of coordination shells 
in the RDF, large MS contributions in EXAFS due to the presence of 
linear --Cu--N--Cu-- chains  and an anisotropy 
of atom vibrations due to tilting motion of NCu$_6$ octahedra. 
Since RMC simulation results in a 3D model of the structure, one has an opportunity 
to analyse separately behaviour of atoms, belonging to different coordination 
shells but located at close distances from the absorber.  
Temperature dependences of the MSRD factors for selected Cu--N and Cu--Cu 
atom pairs were calculated from atomic configurations obtained by RMC and are
shown in Fig.~\ref{fig6} (lower panel). Strong correlation in atomic motion 
was found for atoms (N$_1$, Cu$_{3a}$ and N$_{7a}$) located in the chains along the crystallographic axes. Moreover, it is possible to distinguish clearly 
large difference in the MSRD factors of non-equivalent atoms located in 
the 3rd (Cu$_{3a}$ and Cu$_{3b}$) and  7th  (N$_{7a}$ and N$_{7b}$) shells. 
Strong increase of the MSRD of  Cu$_{3a}$, Cu$_{2}$ and Cu$_{3b}$ points to the anisotropic vibration of copper atoms in the direction orthogonal to 
--N--Cu--N-- chains.

\section{Conclusions}
\label{conc}

Atomistic simulation methods such as molecular dynamics and reverse Monte Carlo 
provide a natural way to include disorder (static and dynamic)
into the EXAFS formalism taking into account multiple-scattering  effects.

The two methods have several common points. 
In both cases, multiple absorption edges can be easily simulated or fitted, 
thus improving the reliability of the accessible structural information.
The analysis of EXAFS contributions from outer 
coordination shells of the absorbing atom is feasible, which is rather challenging in 
conventional approach  but provides an access to some useful structural and dynamic 
properties of a material as, for example, mean-square displacements.

Opposite to conventional analysis, dealing with a set of structural parameters,
MD-EXAFS and RMC-EXAFS approaches provide a result in terms of atomic configurations,
giving information on atom-atom and bond-angle distributions and correlations.  
Moreover, an access to atomic coordinates makes it possible to distinguish  
contributions of non-equivalent atom pairs with equal or close path lengths.

At the same time, there are also several differences between the two methods. 

The MD-EXAFS approach does not require any structural fitting parameters, and 
the structural model of a material is uniquely defined by the results of the 
MD simulation. The agreement between the experimental and calculated 
CA EXAFS spectra depends on the accuracy of interatomic potential model,  
therefore, EXAFS spectrum can be used to validate the interatomic potentials. 
	
3D structure models obtained by the RMC method from experimental EXAFS can be directly 
compared with the results of other atomistic simulations. Moreover, they can be employed 
to include disorder effects into first-principles simulations to 
predict temperature dependent material properties. Note that constraints can be easily 
incorporated into the RMC analysis to account for information from other experiments 
(diffraction, total scattering, etc) or chemical/geometrical information 
(bond-lengths, bonding angles, coordination, energetics, etc). 

\section*{Acknowledgements}
This work has been partially supported by the Latvian Council of Science project no. lzp-2018/2-0353.

%\bibliography{xafs17}
%\bibliographystyle{elsarticle-harv}

\end{document}